\documentclass[10pt,journal,a4paper]{IEEEtran}
\usepackage[numbers, square, comma, sort&compress]{natbib}

\ifCLASSOPTIONcompsoc
\else
\fi

\ifCLASSINFOpdf
\else
\fi

\usepackage[sf,SF]{subfigure}
\usepackage{psfrag}
\usepackage{pstricks}
\usepackage{epsfig}
\usepackage{amsmath,amsthm}
\usepackage{amssymb, bm, dsfont}
\usepackage{float}
\usepackage{amssymb,stmaryrd,amsmath,amsfonts,rotating}

\providecommand{\Z}{\ensuremath{\mathds{Z}}}

\providecommand{\N}{\ensuremath{\mathds{N}}}

\newcommand{\G}{\mathcal{G}}

\newcommand{\e}{\epsilon}

\newtheorem{lem}{Lemma}
\newtheorem{thm}{Theorem}

\newtheorem{coro}{Corollary}
\theoremstyle{definition}
\floatstyle{ruled} 
\newfloat{algorithm}{tbp}{loa}
\floatname{algorithm}{Algorithm}
\newcommand{\Ldens}[1]{\mathsf{\MakeLowercase{#1}}}

\begin{document}
%
\title{Universal Rateless Codes From Coupled LT Codes}
\author{
	\IEEEauthorblockN{
		Vahid Aref and R{\"u}diger L. Urbanke \\
		\IEEEauthorblockA{EPFL, Lausanne, Switzerland, Email: vahid.aref@epfl.ch, rudiger.urbanke@epfl.ch }\thanks{This work was supported by grant No. 200021-125347 of the Swiss
National Foundation.}}
}
\maketitle

\IEEEcompsoctitleabstractindextext{%
\begin{abstract}
It was recently shown that spatial coupling of individual low-density
parity-check codes improves the belief-propagation threshold of
the coupled ensemble essentially to the maximum a posteriori threshold of the
underlying ensemble. We study the performance of spatially coupled
low-density generator-matrix ensembles when used for transmission over
binary-input memoryless output-symmetric channels. We show by means of
density evolution that the threshold saturation phenomenon also takes
place in this setting.
Our motivation for studying low-density generator-matrix codes is
that they can easily be converted into rateless codes. Although there
are already several classes of excellent rateless codes known to
date, rateless codes constructed via spatial coupling might offer
some additional advantages.  In particular, by the very nature of
the threshold phenomenon one expects that codes constructed on this
principle can be made to be universal, i.e., a single construction can
uniformly approach capacity over the class of binary-input memoryless
output-symmetric channels.
We discuss some necessary conditions on the degree distribution which
universal rateless codes based on the threshold phenomenon have to
fulfill.  We then show by means of density evolution and some simulation
results that indeed codes constructed in this way perform very well over
a whole range of channel types and channel conditions. 
\end{abstract}

}


\IEEEdisplaynotcompsoctitleabstractindextext

%
\IEEEpeerreviewmaketitle

\section{Introduction}

%
%

%
%
%
%
\IEEEPARstart{T}{he} idea of spatially coupling copies of a
graphical model was introduced for the coding context in 
\cite{ZigFel} in the form of convolutional LDPC ensembles.
The performance of such ensembles was investigated in, among others,
\citep{IDLDPCC,TerminLDPCCCthreshold,ProtoLDPCC} and it was found to
be very good. In particular the threshold of a coupled ensemble was
consistently found to be significantly superior to the threshold of the
underlying ensemble.  It was then shown in \citep{couplLDPC,CouplBMS} why
this is the case, and the phenomena was termed {\em threshold saturation.}

The key observation in the above papers is that the  \emph{belief
propagation} (BP) threshold of the coupled ensemble is considerably
improved and becomes close to  the \emph{maximum a posteriori} (MAP)
threshold of the underlying ensemble while the MAP threshold of
the coupled and underlying ensembles are close to each other.
This phenomenon has also been observed in several other classes of
graphical models~\citep{couplHassani,HNR} and seems to be rather general:
when we spatially couple, the dynamical threshold of the chain converges
to the static threshold of the un-coupled model.

We study the coupling phenomenon for low-density generator-matrix (LDGM)
codes. LDGM codes are closely related to LT codes. LT codes
were originally designed for communication over the binary erasure
channels (BEC) with unknown erasure probability~\cite{Luby02ltcodes}. For
these codes the encoder generates an (in principle) infinite sequence of
output symbols. The decoder collects as many output symbols as necessary
to successfully recover all the information bits. LT codes are one
of the first instances of \emph{rateless} codes, see \cite{RapBSCetesami}.
They are called rateless codes because the rate of the code is not
fixed a priori and can vary from essentially zero to essentially one,
depending on the channel condition.

A typical application of rateless codes is a system where the actual
channel is unknown to the encoder and chosen from a given uncertainty
set. LT codes can asymptotically reach $1-\mu$ of the capacity of the BEC
with unknown erasure probability, for any $\mu>0$~\cite{Luby02ltcodes}. In
particular, LT codes are \emph{universal} over the BEC.  By adding a
proper precoder to the LT codes, Shokrollahi introduced \emph{Raptor
codes} which exhibit an even better performance in terms of
encoding/decoding complexity and error probability~\cite{Raptor}.
There is a considerable literature on rateless codes. Let us just
mention a very small selection and refer the reader to some of the review
articles for a more thorough literature review.  The error performance of
Raptor codes and LT codes over binary-input memoryless output-symmetric
(BMS) channels was investigated in \cite{ratelessYedidia}.  Later in
\cite{RapBSCetesami}, the authors showed how to design $\mu$-capacity
achieving Raptor codes, for arbitrary $\mu>0$, on the binary symmetric
channel (BSC) and the binary additive white Gaussian noise channel
(BAWGNC); the authors also proved that LT codes are not universal
over the BSC and the BAWGN channel families.  

The objective of this paper is to introduce a further alternative to the
construction of rateless codes, namely to construct rateless codes
via spatial coupling of LT ensembles. We show by means of density
evolution that the threshold saturation also takes place in this setting.
We provide some necessary conditions on the degree distributions in
order for the constructed ensemble to be universal.

We describe the structure of coupled ensembles in
Section~\ref{sec:structure}.  There we also explain the relationship
between LT and LDGM ensembles.  The saturation phenomenon is investigated
in Section~\ref{sec:saturation}. We derive some necessary conditions
for such an ensemble to be universal in Section~\ref{sec:discussion}. We
also provide some simulation results for various channel types and rates
which give further support for our conjecture.

\section{Rateless Ensembles from Coupled LT Ensembles}\label{sec:structure} 
We propose to construct rateless codes by spatially coupling LT codes.
When the number of information bits and the number of output bits tends
to infinity (at a fixed ratio), the performance of such a structure can in
turn be assessed by analyzing an ensemble of spatially coupled LDGM codes.
Let us start by recalling the definition of LT codes.

\subsubsection{Structure of LT Ensemble}
Let $u_1, \dots, u_m$ denote the information bits we want to transmit.
For LT codes, in principle, an infinite stream of output symbols is
generated from these $m$ information bits.  The receiver ``listens'' to as
many of them as needed in order to decode the $m$ information bits reliably.

More precisely, the encoder generates a sequence of output symbols as
follows: First, an integer $d$, called the degree, is independently
and randomly chosen according to a given degree distribution. This
distribution is encoded by the polynomial $R(x)= \sum_{d=1}^{d_{\max}}
R_d x^d$, where $R_d$ is the probability of choosing $d$.  Next, a
$d$-tuple of information bits is uniformly picked from all $m\choose d$
distinct $d$-tuples, denote it by $(i_1, \dots, i_d)$.  Finally, the sum
$u_{i_1}+\cdots+u_{i_d}$ is computed (also called the ``output symbol'')
and it is transmitted over the channel.  Here, we assumed that the
transmitter and the receiver share randomness so that the choice of
the degree as well as the choice of the indices is known both to the
transmitted and to the receiver.

The receiver collects a number of output symbols (typically at least
equal to the number of information bits) and starts the decoding process
using the BP algorithm. If it cannot decode given this information, it
collects further output symbols and retries. It continues in this manner
until all $m$ information bits are decoded.  Assume that the receiver
decodes all information bits using $n$ output symbols. We then say that
the code has rate $r={m\over n}$.

The received output symbols and information bits can be represented by a
bipartite graph $\G(U,G;E)$. Here $U$ denotes the set of information nodes
and it has cardinality $m$. In the same manner, $G$ denotes the set of
\emph{generator nodes} (output symbols). The set $E$ denotes all edges;
there is an edge between a generator node and an information node iff
the corresponding bit was used in the computation of this output symbol.

\subsubsection{Coupled LT Ensemble}
\begin{figure}[tb]
\centering
\begin{picture}(200,120)(0,0)
{
\put(0,0){\includegraphics[width=0.8\linewidth]{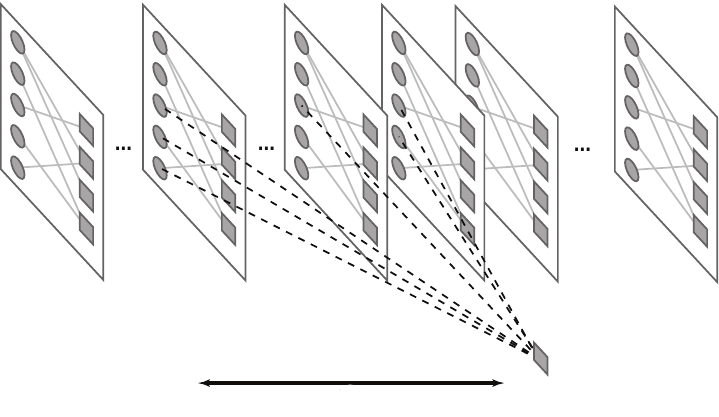}}
\put(20,10){\makebox(0,0)[rb]{$0$}}
\put(220,10){\makebox(0,0)[rb]{$L-1$}}
\put(100,-3){\makebox(0,0)[rb]{$w$}}
}
\end{picture}
\caption{\label{fig:CoupleLDGM} Adding a new generator node to a coupled LDGM ensemble.}
\end{figure}
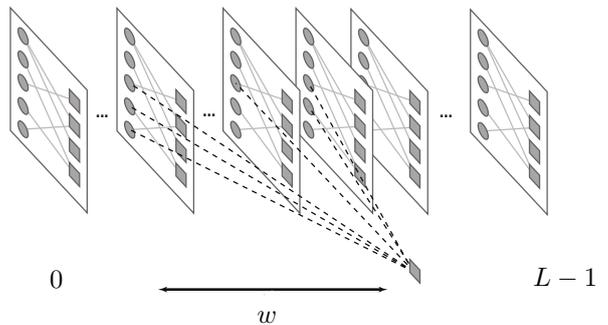

Let us now discuss how to couple LT codes.  Assume that the information
bits are divided into $L$ sets located at positions $[0,L-1]$ and each
having $m$ information bits. Let these bits be labeled from $1$ to
$mL$. Let the generator nodes be located at positions $[0,L+w-2]$, where
$w$ is a \emph{smoothing} parameter, $w \in \N$. To generate an output
symbol, the encoder picks $i\in[0,L+w-2]$. This is the position of the
next generator node which is being constructed.  Next the degree $d$ is
chosen as in the uncoupled case, according to the distribution $R(x)$, and
independently from all previous choices. Then, each of the $d$ connections
is uniformly and independently chosen among the $mw$ information bits
in the range $[i-w+1,i]$, see Fig.~\ref{fig:CoupleLDGM}. For generator
nodes situated close to the boundaries, if the position of a chosen
information bit is not in the range $[0,L-1]$ then the associated edge is
omitted. Equivalently we can assume that it is connected to a bit outside
of this range which is known both to the transmitter and the receiver and whose
value can without loss of generality be assumed to be $0$.  Finally, the
encoder sends the sum of the values of the connected information bits. As
for the uncoupled case, we assume that shared randomness is available at
the transmitter and the receiver so that the choice of positions, degrees,
and connected bits is known on both sides. We call the resulting ensemble
a \emph{spatially coupled LT ensemble}.

\subsubsection{LDGM Ensembles as Limits of LT Ensembles}
Consider an uncoupled LT code.
Since the degree of every generator node is chosen independently according
to the distribution $R(x)$, the empirical distribution of the degrees
of the output symbols converges a.s. to $R(x)$.  Further, if we let
$m$ and $n$ tend to infinity but fix their ratio, then the empirical
degree distribution of the information bits converges a.s. to the Poisson
distribution $\lambda(x)=e^{l_{\text{avg}}(x-1)}$, where $l_{\text{avg}} =
{n\over m} R'(1)$ is the average degree of an information bit.  Therefore,
in this sense (for increasing blocklengths) the resulting code tends to
an instance of the LDGM $(e^{l_{\text{avg}}(x-1)},R(x))$ ensemble. Note also that
for any fixed number of iterations density evolution is continuous in
the degree distribution.  

In order to study the threshold behavior of LT codes we can therefore
study the threshold behavior of the equivalent LDGM ensemble.  Only when
we are interested in the finite-length scaling behavior do we need to
take the small deviations of the degree distribution from the expected
value into account. For coupled LT ensembles the same argument applies.
Therefore, for the purpose of analysis, we consider $L$ copies of
$(e^{l_{\text{avg}}(x-1)},R(x))$ LDGM ensembles spatially coupled in the
same way as described above. 

\subsubsection{Design Rate}
In the coupled setting, the design rate of the code is equal to the
total number of non-trivial information bits, which is equal to $mL$, 
divided by the number of generator bits that are connected to at least 
one of the $mL$ non-trivial information bits. We have,  
 
\begin{lem}[Design Rate]
\label{lem:rateloss}
Consider an $(\lambda(x),R(x),L,w)$ coupled LDGM ensemble such that the
underlying ensemble has $n$ generator nodes and $m$ information bits. The design rate is
$r=\frac{mL}{n (L - w + 1) + 2n\sum_{i=1}^{w-1} (1 - R({i\over w}))}$.
\end{lem}
We see that the design rate of the
coupled ensemble is slightly decreased compared to the design rate
$m/n$ of the underlying ensemble.  However, this
rate loss vanishes with $L$ at a speed of $\Theta({w\over L})$.  Hence,
we should not pick $L$ too small in order to keep the rate loss at an
acceptable level. On the other hand, picking $L$ very large leads to
very long codes. Hence, there is an inherent trade-off. For LDPC
ensembles, various ways of reducing the rate loss were suggested in
\cite{CouplBMS}. The same basic ideas can be applied in the present
setting to substantially mitigate the rate loss.  We will not pursue this
topic further, although in a real setting it is important.

\section{Threshold Saturation of Coupled LDGM Ensemble}\label{sec:saturation}
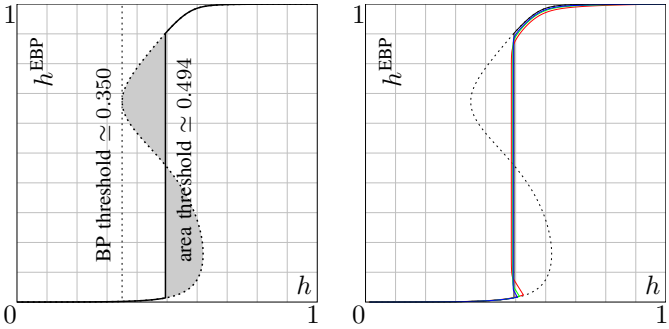
\begin{figure}[tb]
\setlength{\unitlength}{1.0bp}%
\centering
\begin{picture}(160,110)(0,0)
{
\put(-40,0){\rotatebox{0}{\includegraphics[scale=0.7]{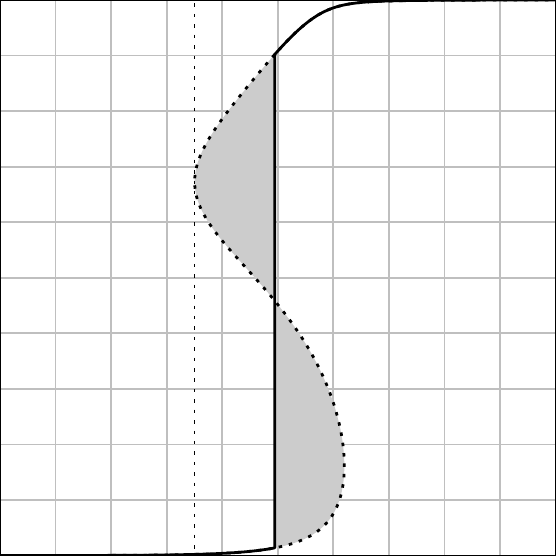}}}
\put(-35,100){\makebox(0,0)[lt]{\rotatebox{90}{$h^{\text{\scriptsize EBP}}$}}}
\put(70,3){\makebox(0,0)[rb]{$h$}}
\put(73,-8){\makebox(0,0)[rb]{$1$}}
\put(-40,105){\makebox(0,0)[rb]{$1$}}
\put(-40,-8){\makebox(0,0)[rb]{$0$}}

\put(90,0){\rotatebox{0}{\includegraphics[scale=0.7]{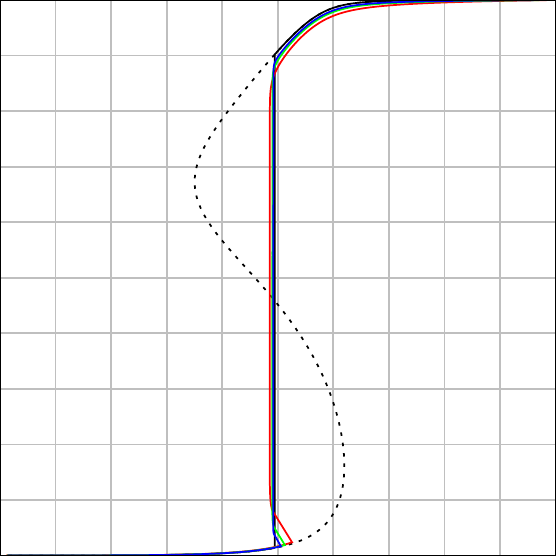}}}
\put(95,100){\makebox(0,0)[lt]{\rotatebox{90}{$h^{\text{\scriptsize EBP}}$}}}
\put(200,3){\makebox(0,0)[rb]{$h$}}
\put(203,-8){\makebox(0,0)[rb]{$1$}}
\put(90,105){\makebox(0,0)[rb]{$1$}}
\put(90,-8){\makebox(0,0)[rb]{$0$}}
{
\footnotesize
\put(-5,15){\makebox(0,0)[rb]{\rotatebox{90}{BP threshold $\simeq 0.350$}}}
\put(25,15){\makebox(0,0)[rb]{\rotatebox{90}{area threshold $\simeq 0.494$}}}

}
}
\end{picture}
\caption{\label{fig:coupledsystem} Left: EBP GEXIT curve (dashed) 
and Maxwell curve (solid) of the $(e^{12.32(x-1)},R_1(x))$ LDGM
ensemble with $R_1(x)=0.02 x + 0.6 x^2 + 0.38 x^{13}$ and  $r=0.5$
for transmission over the BEC. The area under the Maxwell curve is $r$
and the area threshold is $0.494$. The BP threshold is $0.350$. Right:
The EBP GEXIT curves of the corresponding coupled ensembles for $(L,w)$
equal to $(64,5)$ (red curve), $(128,5)$ (green curve), and $(512,11)$
(blue curve). Note that these EBP GEXIT curves are all very close to
the Maxwell curve of the underlying ensemble.} \end{figure}

Let us consider as example the EBP GEXIT curve of the $(e^{12.32(x-1)},R_1(x))$
LDGM ensemble where $R_1(x)= 0.02 x + 0.6 x^2 + 0.38 x^{13}$ and
$r=\frac{1}{2}$.\footnote{In a nutshell, the EBP GEXIT curve is the curve
of all fixed points (FP) of density evolution (DE) for the given ensemble.
For an in-depth discussion we refer the reader to \cite{URbMCT}.}  In
the left picture in Fig.~\ref{fig:coupledsystem} the EBP GEXIT curve
is shown as a dashed line. For the given example it has the shape of an
``S.''  Let us construct the {\em Maxwell} curve. We get the Maxwell
curve by taking the EBP GEXIT curve and by cutting the ``S'' by means of a
vertical line, where the line is located in such a way that the two gray
areas are equal (see the left picture). The Maxwell curve then consists
of the the vertical line plus the two connecting parts of the EBP GEXIT
curve so that the total curve represents an increasing function.  In the
sequel we will refer to the entropy value where the vertical line is
located as the \emph{area threshold} (since the position of the vertical
line is defined by an equality of areas). In Fig.~\ref{fig:coupledsystem}
the Maxwell curve is shown as solid black curve and the area threshold is
$h \approx 0.494$.  This has to be compared to the BP threshold of this
ensemble which can be seen to be around $0.35$. The significance of the
Maxwell curve is that for a wide range of ensembles and channels it is
conjectured to characterize the performance of the MAP decoder, see e.g.,
\cite{URbMCT}.
\footnote{Note that, even if we assume that the Maxwell curve characterizes the
MAP performance, the area threshold defined above is not really the
MAP threshold since there is an error floor (this code does not have a
non-trivial MAP threshold). But if we assume, as it is e.g. the case for
Raptor codes, that we are using a pre-code and that the error floor is
sufficiently small, then this area threshold has an important operational
significance. As a consequence of the error floor, this area threshold
can even be slightly larger than the Shannon threshold.}

Let us now show by means of DE computations that the BP
threshold of the coupled ensemble is very close to the area threshold
of the underlying ensemble for a wide range of BMS channels (see
Fig.~\ref{fig:coupledsystem}). This observation suggests that the
threshold saturation phenomenon also occurs in the current setting.

\begin{figure*}[htb]
\setlength{\unitlength}{0.6bp}%
\centering
\begin{picture}(520,510)(0,0)
{
\footnotesize
\put(340,0){\rotatebox{0}{\includegraphics[scale=0.6]{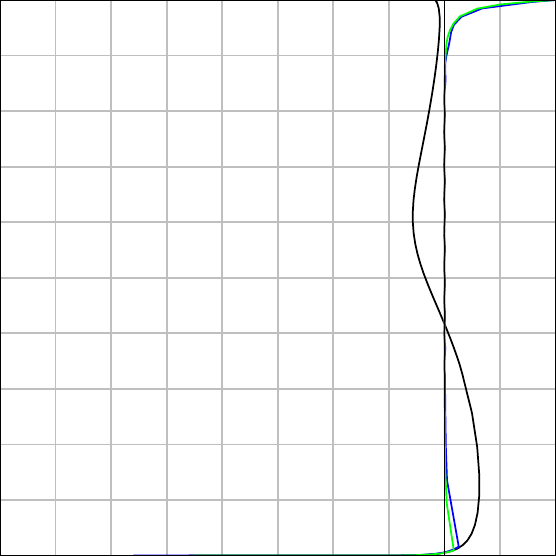}}}
\put(440,140){\makebox(0,0)[rt]{$\text{\scriptsize AWGN}$}}
\put(345,140){\makebox(0,0)[lt]{\rotatebox{90}{$h^{\text{\scriptsize EBP}}$}}}
\put(495,5){\makebox(0,0)[rb]{$h$}}
\put(500,-10){\makebox(0,0)[rb]{$1$}}
\put(339,160){\makebox(0,0)[rt]{$1$}}
\put(341,-10){\makebox(0,0)[rb]{$0$}}
{
\footnotesize
\put(465,15){\makebox(0,0)[rb]{\rotatebox{90}{$h^{\text{\tiny \text{Sha}}}{\text{\small = 0.8}}$}}}
\
}
\footnotesize
\put(170,0){\rotatebox{0}{\includegraphics[scale=0.6]{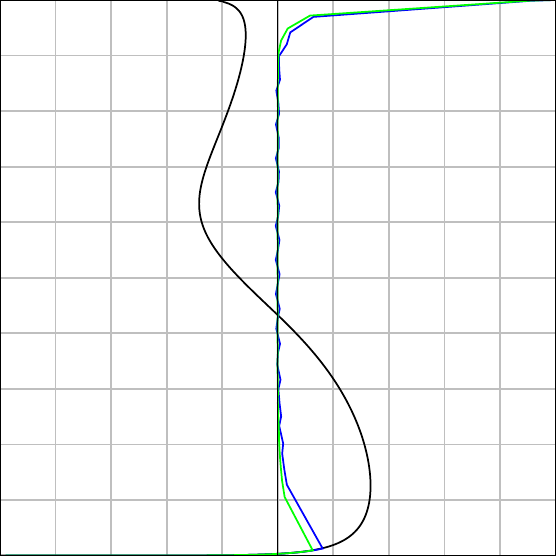}}}
\put(310,140){\makebox(0,0)[rt]{$\text{\scriptsize AWGN}$}}
\put(175,140){\makebox(0,0)[lt]{\rotatebox{90}{$h^{\text{\scriptsize EBP}}$}}}
\put(325,5){\makebox(0,0)[rb]{$h$}}
\put(330,-10){\makebox(0,0)[rb]{$1$}}
\put(169,160){\makebox(0,0)[rt]{$1$}}
\put(171,-10){\makebox(0,0)[rb]{$0$}}
{
\footnotesize
\put(245,15){\makebox(0,0)[rb]{\rotatebox{90}{$h^{\text{\tiny \text{Sha}}}{\text{\small = 0.5}}$}}}
\
}
\footnotesize
\put(0,0){\rotatebox{0}{\includegraphics[scale=0.6]{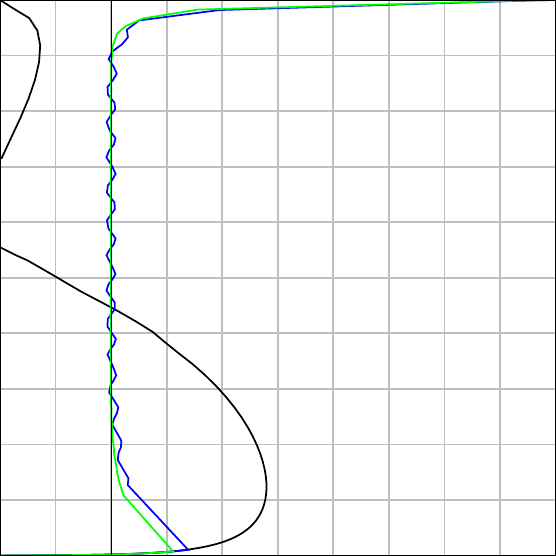}}}
\put(140,140){\makebox(0,0)[rt]{$\text{\scriptsize AWGN}$}}
\put(5,140){\makebox(0,0)[lt]{\rotatebox{90}{$h^{\text{\scriptsize EBP}}$}}}
\put(155,5){\makebox(0,0)[rb]{$h$}}
\put(160,-10){\makebox(0,0)[rb]{$1$}}
\put(-1,160){\makebox(0,0)[rt]{$1$}}
\put(1,-10){\makebox(0,0)[rb]{$0$}}
{
\footnotesize
\put(30,15){\makebox(0,0)[rb]{\rotatebox{90}{$h^{\text{\tiny \text{Sha}}}{\text{\small = 0.2}}$}}}
\
}
}
{
\footnotesize
\put(340,175){\rotatebox{0}{\includegraphics[scale=0.6]{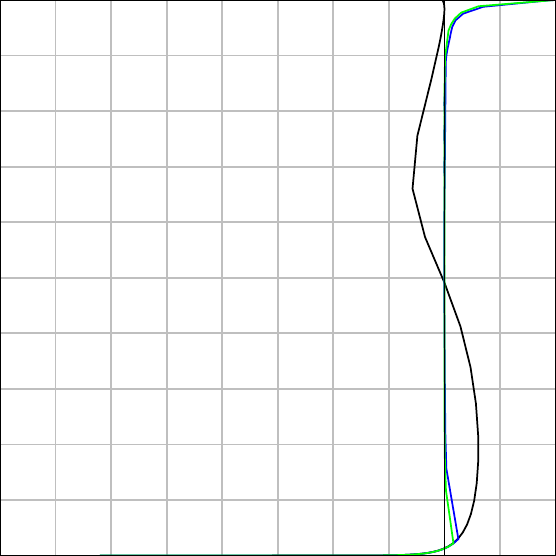}}}
\put(440,315){\makebox(0,0)[rt]{$\text{\scriptsize BSC}$}}
\put(345,315){\makebox(0,0)[lt]{\rotatebox{90}{$h^{\text{\scriptsize EBP}}$}}}
\put(495,180){\makebox(0,0)[rb]{$h$}}
\put(500,165){\makebox(0,0)[rb]{$1$}}
\put(339,335){\makebox(0,0)[rt]{$1$}}
\put(341,165){\makebox(0,0)[rb]{$0$}}
{
\footnotesize
\put(465,190){\makebox(0,0)[rb]{\rotatebox{90}{$h^{\text{\tiny \text{Sha}}}{\text{\small = 0.8}}$}}}
\
}
\footnotesize
\put(170,175){\rotatebox{0}{\includegraphics[scale=0.6]{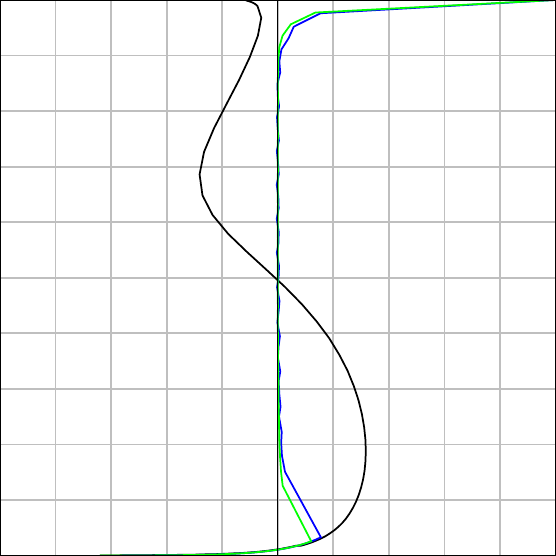}}}
\put(310,315){\makebox(0,0)[rt]{$\text{\scriptsize BSC}$}}
\put(175,315){\makebox(0,0)[lt]{\rotatebox{90}{$h^{\text{\scriptsize EBP}}$}}}
\put(325,180){\makebox(0,0)[rb]{$h$}}
\put(330,165){\makebox(0,0)[rb]{$1$}}
\put(169,335){\makebox(0,0)[rt]{$1$}}
\put(171,165){\makebox(0,0)[rb]{$0$}}
{
\footnotesize
\put(245,190){\makebox(0,0)[rb]{\rotatebox{90}{$h^{\text{\tiny \text{Sha}}}{\text{\small = 0.5}}$}}}
\
}
\footnotesize
\put(0,175){\rotatebox{0}{\includegraphics[scale=0.6]{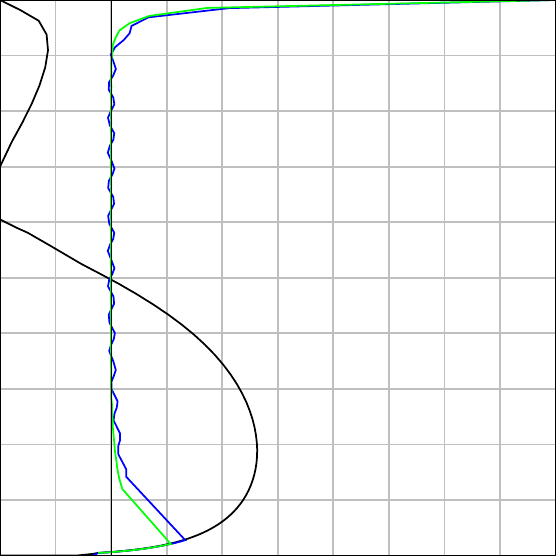}}}
\put(140,315){\makebox(0,0)[rt]{$\text{\scriptsize BSC}$}}
\put(5,315){\makebox(0,0)[lt]{\rotatebox{90}{$h^{\text{\scriptsize EBP}}$}}}
\put(155,180){\makebox(0,0)[rb]{$h$}}
\put(160,165){\makebox(0,0)[rb]{$1$}}
\put(-1,335){\makebox(0,0)[rt]{$1$}}
\put(1,165){\makebox(0,0)[rb]{$0$}}
{
\footnotesize
\put(30,190){\makebox(0,0)[rb]{\rotatebox{90}{$h^{\text{\tiny \text{Sha}}}{\text{\small = 0.2}}$}}}
\
}
}

{
\footnotesize
\put(340,350){\rotatebox{0}{\includegraphics[scale=0.6]{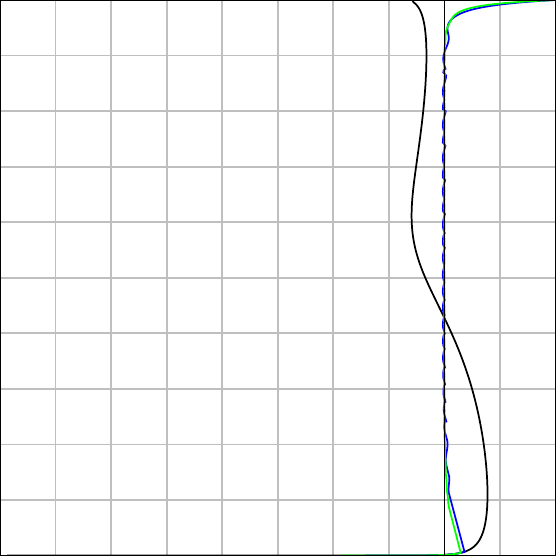}}}
\put(440,490){\makebox(0,0)[rt]{$\text{\scriptsize BEC}$}}
\put(345,490){\makebox(0,0)[lt]{\rotatebox{90}{$h^{\text{\scriptsize EBP}}$}}}
\put(495,355){\makebox(0,0)[rb]{$h$}}
\put(500,340){\makebox(0,0)[rb]{$1$}}
\put(339,510){\makebox(0,0)[rt]{$1$}}
\put(341,340){\makebox(0,0)[rb]{$0$}}
{
\footnotesize
\put(465,365){\makebox(0,0)[rb]{\rotatebox{90}{$h^{\text{\tiny \text{Sha}}}{\text{\small = 0.8}}$}}}
\
}
\footnotesize
\put(170,350){\rotatebox{0}{\includegraphics[scale=0.6]{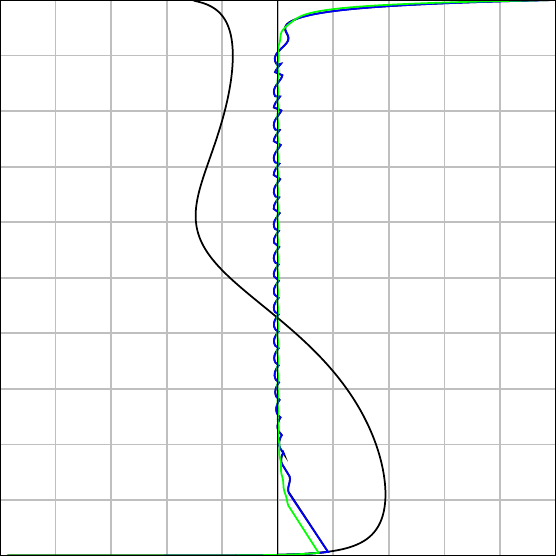}}}
\put(310,490){\makebox(0,0)[rt]{$\text{\scriptsize BEC}$}}
\put(175,490){\makebox(0,0)[lt]{\rotatebox{90}{$h^{\text{\scriptsize EBP}}$}}}
\put(325,355){\makebox(0,0)[rb]{$h$}}
\put(330,340){\makebox(0,0)[rb]{$1$}}
\put(169,510){\makebox(0,0)[rt]{$1$}}
\put(171,340){\makebox(0,0)[rb]{$0$}}
{
\footnotesize
\put(245,365){\makebox(0,0)[rb]{\rotatebox{90}{$h^{\text{\tiny \text{Sha}}}{\text{\small = 0.5}}$}}}
\
}

\footnotesize
\put(0,350){\rotatebox{0}{\includegraphics[scale=0.6]{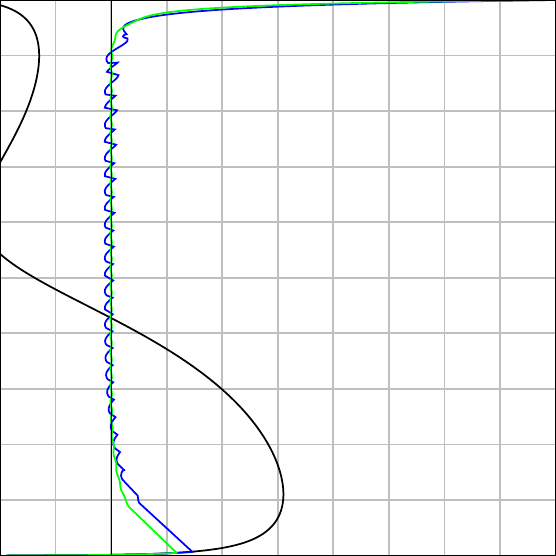}}}
\put(140,490){\makebox(0,0)[rt]{$\text{\scriptsize BEC}$}}
\put(5,490){\makebox(0,0)[lt]{\rotatebox{90}{$h^{\text{\scriptsize EBP}}$}}}
\put(155,355){\makebox(0,0)[rb]{$h$}}
\put(160,340){\makebox(0,0)[rb]{$1$}}
\put(-1,510){\makebox(0,0)[rt]{$1$}}
\put(1,340){\makebox(0,0)[rb]{$0$}}
{
\footnotesize
\put(30,365){\makebox(0,0)[rb]{\rotatebox{90}{$h^{\text{\tiny \text{Sha}}}{\text{\small = 0.2}}$}}}
\
}
}
\end{picture}
\caption{\label{fig:GEXIT} EBP GEXIT curves of LDGM ensemble
$(\lambda(x),R_2(x))$ (black curves) and the corresponding coupled ensembles for
$(L,w)$ equal to $(32,3)$ (blue curves) and $(64,4)$ (green curves)
where $R_2(x)=0.360 x^2 + 0.313 x^3 + 0.327 x^{22}$. The
ensembles are depicted for different rates 0.2, 0.5 and 0.8 in BEC,
BSC and AWGN channel. For the rate 0.8, there is no BP threshold for
the underlying LDGM ensembles. For all cases, the area threshold of
individual ensembles is very close to the
Shannon threshold.} \end{figure*}
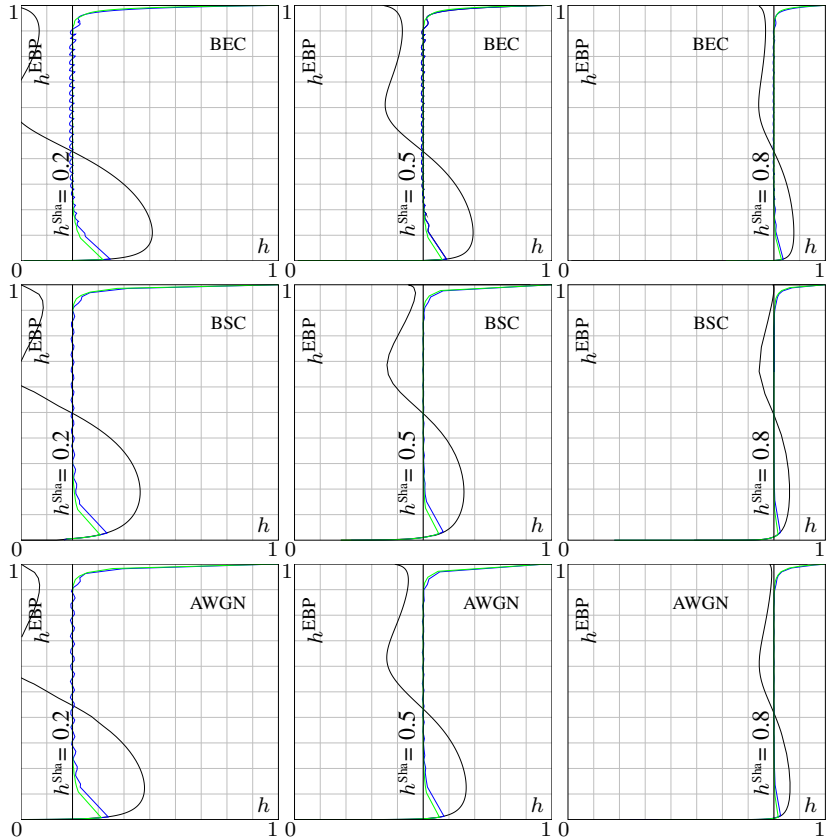

The first step in the evaluation of the asymptotic performance is to write
down the density evolution (DE) equations. To keep the notation at a manageable
level, let us start with the case of the BEC.
\begin{lem}[DE Equations]
\label{lem:DEupdate}
Consider a coupled $(\lambda(x)=e^{l_{\text{avg}}(x-1)},R(x),L,w)$ LDGM
ensemble and transmission over the BEC with erasure probability $\e$.
Let $x_i,\; i\in \Z$, denote the average erasure probability which
is emitted by information nodes at position $i$ and $y_j,\; j\in \Z$,
denote the average erasure probability which is emitted by generator
nodes at position $j$.  The fixed point (FP) condition implied by DE is then
\begin{align}
\label{eq:xupdate}
x_i = \lambda({1\over w}\sum_{j=i}^{i+w-1} y_j), \\ 
\label{eq:yupdate}
y_j = 1 - (1-\e)\rho(1-{1\over w}\sum_{i=j-w+1}^{j} x_i), 
\end{align}
where $\rho(x)= R'(x) / R'(1)$ and $x_i=0$ for $i\not\in [0,L-1]$.
\end{lem}

As mentioned before, since the information bits outside the interval
$[0,L-1]$ are known, we can assume that $x_i=0$ for $i\not\in [0,L-1]$.
The decoding process starts with $x^{(0)}_i=1$ for $i\in
[0,L-1]$.  Let $x^{(l)}_i$ denote the average erasure probability
which is emitted by information bits at position $i$ at round $l$.
If at each decoding round all $x^{(l)}_i$ are updated according to
\eqref{eq:xupdate} and \eqref{eq:yupdate}, then for each $i$
the sequence $x^{(l)}_i$ is monotonically decreasing.
Since the sequence is bounded from below it must converge.
Call the limit $x^{(\infty)}_i$, $i\in
[0,L-1]$. We call the vector $(x^{(\infty)}_0,\cdots,x^{(\infty)}_{L-1})$ the
forward DE FP for the erasure probability $\e$. From this
we can compute the BP GEXIT value $h_j$ at the generator nodes in position $j\in
[0,L+w-2]$. It is defined by,
$h_j(x_0,\cdots ,x_{L-1})=1-R(1-{1\over w}\sum_{i=j-w+1}^{j} x_i)$.
The same analysis can be performed for general BMS channels
by writing down the corresponding DE equations. Since this 
is rather routine, we skip this part.


Let us now illustrate the results of the DE analysis. Consider the
LDGM ensemble in Fig.~\ref{fig:coupledsystem}. As discussed above, the
left picture shows its EBP GEXIT curve as well as the derived Maxwell
curve. The right picture shows the EBP GEXIT curves of the corresponding
coupled ensembles with various values of $L$ and $w$.  As we can see
from the picture, all these curves are very close to the Maxwell curve
of the underlying ensemble and seem to approach the closer the larger
we choose $L$ and $w$ (as long as $w$ is small compared to $L$).  So,
according to this numerical evidence, the threshold saturation phenomenon
occurs for this ensemble as conjectured.

Fig.~\ref{fig:GEXIT} shows the EBP GEXIT curves for several further
examples.  All pictures are for the degree distribution $R_2(x) = 
0.360 x^2 + 0.313 x^3 + 0.327 x^{22}$ and $(L,w)$ equal to $(32,3)$
and $(64, 4)$.  The rows correspond to transmission over the BEC, BSC,
and BAWGNC (top to bottom), respectively.  The columns corresponds to
rates $0.2$, $05$, and $0.8$ (left to right), respectively.  For all
these cases we see that the threshold saturation phenomenon takes place.
Also, the resulting thresholds are all very close to the Shannon capacity.
Indeed, we see that this code is uniformly good over these classes of
channels and a wide range of rates. This gives further evidence to our
conjecture that rateless codes constructed on coupling of LT codes can
be made to be universal.


\section{Necessary Conditions on Degree Distribution for Universality
of Coupled LDGM ensemble}\label{sec:discussion}
Although currently we do not know how to prove that coupled LT ensembles
can be made universal, it is easy to derive some necessary conditions
for this to happen.
\begin{itemize}
\item[(i)]{\em Error Floor}: Since we are dealing essentially with
LDGM ensembles, our construction has generically a bit error floor. To
achieve capacity, we have to ensure that this error floor tends to zero
as the block-length grows large. This induces a constraint on average
degree of the generator nodes, see \cite{Raptor}.\footnote{Here we assume that we want
to construct our sequence of capacity achieving ensembles in such a way
that the rate of the outer code tends to one.}

\item[(ii)]{\em Threshold Behavior}:
The premise of coupled ensembles is that their BP threshold is equal to
their area threshold. 

Assuming the above premise, for a given generator degree distribution
and a specific design rate, the corresponding coupled LDGM ensemble is
therefore asymptotically (when $L$ and $w$ tend to infinity) capacity
achieving for a family of channels if the area threshold of the underlying
LDGM ensemble is equal to its Shannon threshold.  If this property holds
for any design rate, then we say that the coupled ensemble with that
degree distribution is universal on that family of channels. If it holds
in addition for any channel family (within lets say the BMS channel family)
then we say that the ensemble is universally capacity achieving.

We will see that this induces a constraint on $R_1$ and $R_2$.
\end{itemize}

\subsection{Error Floor}
LDGM ensembles have in general a non-zero bit error probability below
the ``threshold'' (which we call the error floor) and this error floor
remains essentially unchanged by coupling. 
\begin{thm}[Lower Bound on Error Floor]
The error floor of the $(\lambda(x)=e^{l_\text{avg}(x-1)},R(x))$ LDGM ensemble
when transmission takes place over the BEC with erasure probability $\e$ is lower bounded by
\begin{equation}\label{eq:errorfloor}
P_e \geq \frac{1}{2}\lambda(\e)(1+ l_\text{avg} (1-\e) (1-\frac{R'(1-\lambda(\e))}{R'(1)})).
\end{equation}
Hence, a necessary condition for this expression to tend to zero at a fixed
erasure probability $\e$ is that $R'(1)$ tends to infinity.
\end{thm}

\subsection{Threshold Behavior}
It is shown in \cite{RapBSCetesami} that in order for a
$(e^{l_{\text{avg}}(x-1)},R(x))$ LDGM ensemble (the asymptotic LT code
with degree distribution $R(x)$ in their work) to be capacity achieving
under BP decoding for a BMS channel with LLR density $\Ldens{a}$, the
following two conditions must be fulfilled:
\begin{itemize}
\item[(i)] $R_1 = 0$,
\item[(ii)]$R_2 = \frac{C(\Ldens{a})}{2D(\Ldens{a})}$, where $C(\Ldens{a})$ denoted the capacity of the channel and $D(\Ldens{a}) =\int_{-\infty}^{\infty} a(l) \tanh(l/2) \text{d}l$.
\end{itemize} 
Let us quickly review why these conditions are necessary.  The first
condition is due to the fact that if $R_1>0$, the probability that an
information bit is connected to more than one generator node with degree
one is strictly positive. Imagine that e.g. two generator nodes of degree
one are connected to the same information node. With positive probability
both of them are received. But then clearly one of the generator nodes
is redundant. Hence we are bounded away from capacity.

To explain the second condition we will not follow the arguments used
in \cite{RapBSCetesami} but rather use the language of EBP GEXIT curves.
Assume that $R_1$ is very close
to zero. Let $H(\Ldens{a})=1-C(\Ldens{a})$ be the entropy associated to
the channel with density $\Ldens{a}$. The stability condition of
LDGM ensembles~\cite{URbMCT} implies that the entropy value where
the EBP GEXIT curve deviates from $1$ occurs at the point $\hat
h=H(\hat{\Ldens{a}})$ where $D(\hat{\Ldens{a}}) = \frac{r}{2R_2}$.
Here $r$ is the design rate. 

If $\hat h > 1-r$, then the value of the EBP GEXIT curve at
the Shannon threshold, $1-r$, is strictly smaller than one (see
e.g.~Fig.~\ref{fig:coupledsystem}). Consequently, by applying the
area theorem, the area threshold (and hence the BP threshold) must
be strictly below the Shannon threshold. Therefore, to achieve
capacity, we need that $\hat h \leq 1-r$. This implies that $R_2 \leq
\frac{C(\Ldens{a})}{2D(\Ldens{a})}$.

If $\hat h < 1-r$, then the EBP GEXIT curve deviates from $1$ at $\hat
h$, which lies strictly below the Shannon threshold.  Since the BP
threshold cannot be greater than $\hat h$, also in this case we cannot
achieve capacity (see e.g.~Fig.~\ref{fig:GEXIT}; recall that currently we
discuss uncoupled ensembles).  Therefore, we must have $\hat h \geq 1-r$.
This implies that $R_2 \geq \frac{C(\Ldens{a})}{2D(\Ldens{a})}$.

Combining these two conditions we see that we need equality for the
uncoupled case. From this point of view it is immediately clear why
in this framework we cannot construct universal codes -- the right-hand
side of the above equality depends on the channel!


Consider now the coupled case. Condition (i) is still necessary. What
about condition (ii)? It is easy to see that if the EBP GEXIT curve
deviates from $1$ to the right of the threshold, then also in the coupled
case we are bounded away from capacity. So the condition $\hat h \leq
1-r$, or equivalently, $R_2 \leq \frac{C(\Ldens{a})}{2D(\Ldens{a})}$
still applies.  But, due to the fact that the performance is now given
by the Maxwell curve associated to the underlying ensembles, we are no
longer bound by the second condition. Therefore, we might hope to find
a value of $R_2$ which fulfills the inequality 
$R_2 \leq \frac{C(\Ldens{a})}{2D(\Ldens{a})}$
for all BMS channels.


\begin{lem}[Minimum of Stability Condition]
Over the class of densities $\Ldens{a}$ associated to BMS channels,
\label{lem:bsc}
$\inf_{\Ldens{a}} \frac{C(\Ldens{a})}{2D(\Ldens{a})} = \frac{1}{4\ln(2)}$,
and the minimum is attained for the BSC with entropy $1$.
\end{lem} 

\begin{coro}[Area Threshold of LDGM ensemble]\label{cor:cor}
In order for an $(e^{l_{\text{avg}}(x-1)},R(x), L, w)$ coupled LDGM
ensemble to be asymptotically universal over all BMS channels, it is
necessary that: $\text{(i)} \; R_1 =0, \;\; \text{(ii)} \; R_2 \leq
\frac{1}{4 \ln(2)}\approx 0.3606$.  \end{coro}

Let us look back at the example in Fig.~\ref{fig:GEXIT}. This degree
distribution was designed according to Corollary~\ref{cor:cor}, i.e.,
we have $R_1 = 0$ and $R_2 \leq 0.3606$. Further, $R'(1)\approx 8.85$.
We can see that the GEXIT values of the underlying ensembles are $1$ for
$h > 1-r$ and for all tested channels.  Due to a rather small value for $R_2$,
the BP threshold of the underlying ensemble is quite small (in particular
for low rates) and so the uncoupled ensemble itself would not be useful.
But as we have seen, the performance for the coupled case is universally
close to the Shannon threshold.

Let us summarize: Coupled LDGM codes have the potential advantage of
being universal.  This is indeed a nice property to have for typical
applications of rateless codes.  Further, since we are only concerned
with the area threshold of the underlying ensemble, there are many more
degrees of freedom in their design and typically only a small degree of
irregularity suffices, making them potentially easier to implement.

To be fair, there is a price to be payed. Due to the coupled structure
and the fact that $L$ has to be chosen reasonably large in order to
avoid a large rate loss, it is difficult to construct codes of very
short length which perform well.

\section*{Acknowledgment}
We would like to thank Shrinivas Kudekar and Hamed Hassani for extensive discussions on this
topic and their many suggestions.

\vspace{0.5 cm}
\bibliographystyle{IEEEtran}
\footnotesize
\bibliography{references}

\end{document}